\documentclass[prl,aps,twocolumn,amsfonts,showpacs,superscriptaddress]{revtex4-1}

\pdfoutput=1

\usepackage{epsfig} 
\usepackage{psfrag}
\usepackage{amsmath}
\usepackage{amssymb}
\usepackage{color}
\usepackage{graphicx}
\usepackage{subfigure}
\usepackage{hyperref}

\begin{document}

\title{Holographic Superfluids and the Dynamics of Symmetry Breaking}
\author{M. J. Bhaseen} 
\affiliation{Cavendish Laboratory, University
  of Cambridge, Cambridge, CB3 OHE, U.K.}  
\author{J.  P. Gauntlett} 
\affiliation{Blackett Laboratory, Imperial College,
  London SW7 2AZ, U.K.}  
\author{B. D. Simons} 
\affiliation{Cavendish Laboratory, University
  of Cambridge, Cambridge, CB3 OHE, U.K.}  
\author{J. Sonner} 
\affiliation{DAMTP, University of Cambridge,
  CB3 OWA, U.K.}  
\author{T. Wiseman} 
\affiliation{Blackett
  Laboratory, Imperial College, London SW7 2AZ, U.K.}

\begin{abstract}
  We explore the far from equilibrium response of a holographic
  superfluid using the AdS/CFT correspondence. We establish the
  dynamical phase diagram corresponding to quantum quenches of the
  order parameter source field. We find three distinct regimes of
  behaviour that are related to the spectrum of black hole
  quasi-normal modes. These correspond to damped oscillations of the
  order parameter, and over-damped approaches to the superfluid and
  normal states.  The presence of three regimes, which includes an
  emergent dynamical temperature scale, is argued to occur more
  generally in time-reversal invariant systems that display continuous
  symmetry breaking.
\end{abstract}

\date{July 2012}


\maketitle

In the last few years there has been a wealth of experimental activity
exploring the non-equilibrium properties of quantum many body
systems. Recent advances include observations of long-lived
oscillations in colliding Bose gases \cite{Kinoshita:QNC}, and the
dynamics of cold atoms following a quantum quench
\cite{Sadler:Spont,Smith:Bosequench}. Non-equilibrium measurements
have also been exploited to reveal the superfluid amplitude mode
\cite{Endres:Higgs}, and to explore pairing in high temperature
superconductors \cite{Mansart:Direct}.  In parallel there has also
been significant theoretical work on low-dimensional strongly
correlated systems,  
where analytical \cite{Calabrese:Quench} and
numerical \cite{Vidal:Efficient,White:tDMRG,Daley:tDMRG} progress is
possible; for a review see Ref.~\cite{Polkovnikov:RMP}.

A notable feature to emerge from the dynamics of the integrable BCS
(Bardeen--Cooper--Schrieffer) Hamiltonian, following an abrupt quench
of the pairing interactions, is a regime of persistent oscillations of
the order parameter
\cite{Barankov:Collective,Yuzbashyan:Relax,Yuzbashyan:Solution,Barankov:Coll,Andreev:Noneqfesh}.
This is accompanied by a transition to a regime of damped oscillations
as the quench strength is increased \cite{Barankov:Sync}. These
integrable results apply in the collisionless regime, for timescales
shorter than the energy relaxation time
\cite{Volkov:Collisionless,Gurarie:Strong}.  In spite of these
achievements, it is challenging to see how such results are modified
at late times in the collision dominated regime. In particular, do the
oscillations and the transition withstand quantum and thermal
fluctuations and departures from integrability?  Related
considerations apply to other integrable systems, and generalizing
non-equilibrium results to more generic situations, including higher
dimensions, is a major open challenge.

In this respect, the AdS/CFT (anti-de Sitter/conformal field theory)
correspondence \cite{Maldacena:AdS,Gubser:GTC,Witten:AdSholo} can
offer valuable insights. It recasts certain strongly interacting
quantum systems, which are large $N$ field theories, in terms of
weakly coupled gravitational models in at least one dimension higher.
This provides access to the quantum dynamics from the classical
gravitational equations, where finite temperatures correspond to black
hole solutions
\cite{Danielsson:1999zt,Giddings:1999zu,Bhattacharyya:2009uu,Das:2010yw,
  Albash:2010mv,Sonner:Hawking}.  The methods are very powerful when
combined with numerical solution of the equations of motion, as they
allow access to the far from equilibrium response over the entire time
evolution
\cite{Chesler:2008hg,Murata:Noneq,Bizon:2011gg,Garfinkle:2011hm,Bantilan:2012vu,Buchel:Thermal}.

In this manuscript we will focus on the 
dynamics of a 
holographic superfluid
\cite{Gubser:Breaking,Hartnoll:Building,Murata:Noneq}
under a spatially homogeneous and isotropic quench. Our primary
aim is to reveal  
three regimes of
non-equilibrium response, including a dynamical transition from
under-damped to over-damped collective oscillations.  
We argue that this transition
will feature in other (holographic and non-holographic) time-reversal
invariant systems that display continuous symmetry breaking.

{\em Model.}--- We consider the simplest representative action for a
holographic superfluid, originally introduced in
Refs.~\cite{Gubser:Breaking,Hartnoll:Building}. The model is defined
in the so-called ``bottom-up'' approach which specifies the action
directly on
the gravitational side, without recourse to microscopic string theory calculations.  
It describes a complex scalar field $\psi$, with
charge $q$ and mass $m$, minimally coupled to electromagnetism and
gravity in 3+1 dimensions:
\begin{equation} 
S=\frac{1}{2\kappa^2}\int  d^4x \sqrt{-g}\Big[R+\frac{6}{\ell^2}-\frac{F^2}{4}-|D\psi|^2-m^2|\psi|^2\Big],
\label{model}
\end{equation}
where $F_{ab}=\partial_aA_b-\partial_bA_a$, $D_a=\partial_a-iqA_a$,
and the radius $\ell$ parameterizes the inverse curvature of AdS
space-time. 

\begin{figure}
\includegraphics[width=2.3in,clip]{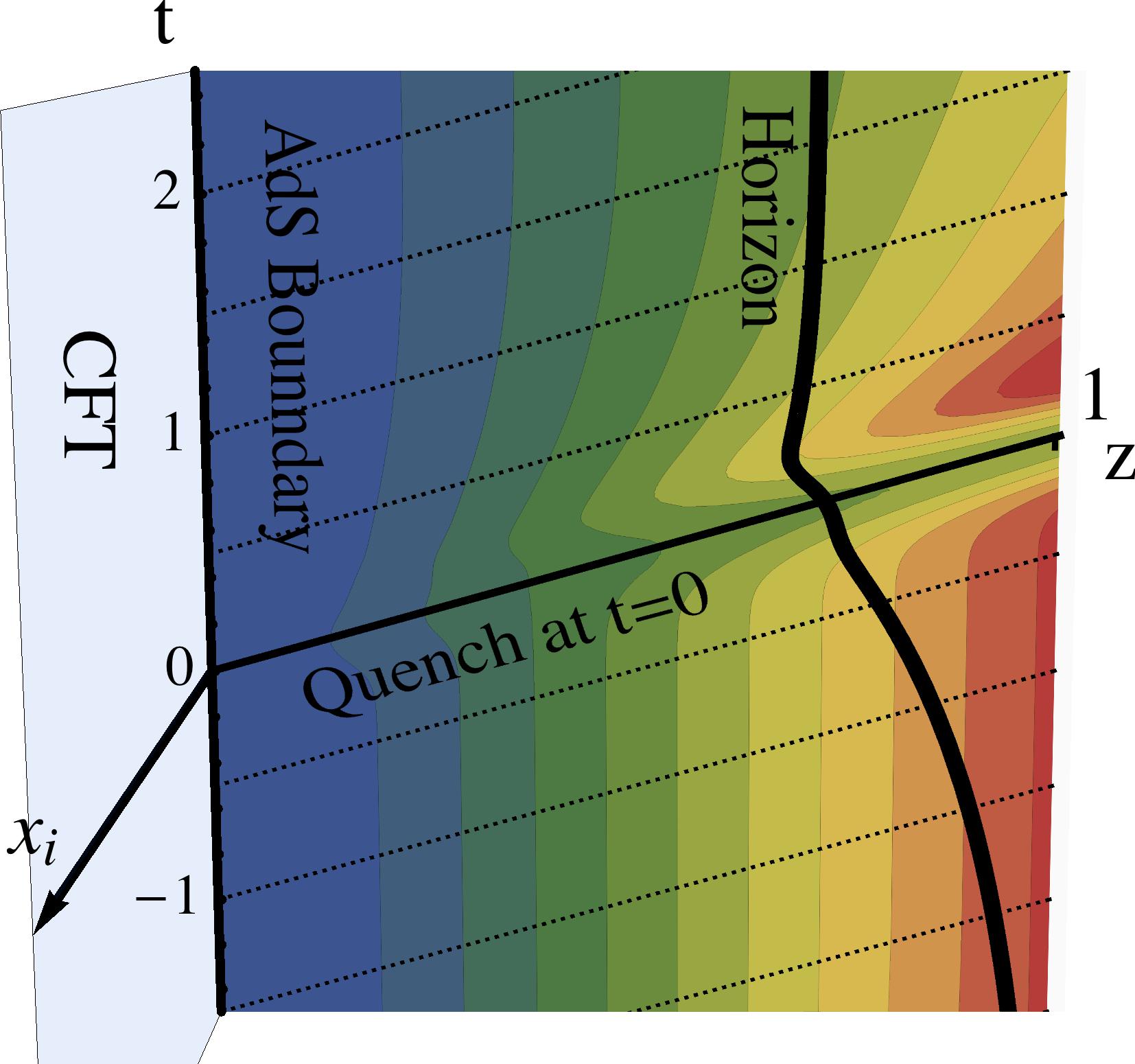}
\vspace*{-0.2cm}
\caption{Schematic representation of the coordinate system; for
    details see the Supplementary Material.  We show data for the
  time evolution of ${\rm Re} \psi(t,z)$ following a Gaussian quench
  at $t=0$ with $\delta=0.15$, from a superfluid black hole initial
  state as $t\to-\infty$ with $T_i/T_c=0.5$.  The behaviour near the
  AdS boundary at $z=0$ is used to extract the dynamics of the
  superfluid order parameter $\langle {\mathcal O}(t)\rangle$ in
  Figs~\ref{Fig:PD},\ref{Fig:Temp}.}
\label{Fig:Setup}
\end{figure}

Exploiting the AdS/CFT correspondence, the model is dual to a strongly
coupled large $N$ CFT in 2+1 D flat Minkowski space-time,
residing on the AdS boundary, as shown in Fig.~\ref{Fig:Setup}; for
reviews see Refs.~\cite{Hartnoll:Lectures,McGreevy:Notes}.  The CFT is
time-reversal invariant and has a global ${\rm U}(1)$ symmetry whose
conserved current, $J_\mu$, is dual to $A_a$.  The ${\rm U}(1)$
symmetry is spontaneously broken below a critical temperature, $T_c$,
corresponding to the onset of superfluidity. This is possible in the
2+1 D CFT due to large $N$ \cite{Hartnoll:Building}.  Other
holographic superfluid models, including 3+1 D CFTs, will exhibit
analogous phenomena. In this manuscript we set
$1/(2\kappa^2)\equiv {\mathcal C}/{\ell}^{2}$ and choose units with
$\ell = 1$.  Here, ${\mathcal C}$ is a measure of the number of local
degrees of freedom in the CFT, with $\mathcal{C} \sim N^{3/2}$ at
large $N$.

In general, it is very difficult to analyse such
strongly interacting high dimensional CFTs, but the AdS/CFT
corrrespondence allows key insights. In particular, fields in AdS
space-time may be related to physical obervables in the CFT via their
coordinate expansion close to the AdS boundary. Assuming spatial
  homogeneity and isotropy of the boundary theory, the holographic description
  requires two coordinates, $z$ and $t$; here $z$ parameterizes the
  distance from the AdS boundary and $t$ is the boundary time, as
  shown in Fig.~\ref{Fig:Setup}.  For example, in equilibrium 
the time-component of
the gauge field, $A_t$, is dual to the charge density $\rho$ of the
CFT, via the expansion $A_t=\mu-z\rho+\dots$, where $\mu$ is the
chemical potential of the CFT with $\langle
J_t\rangle=\rho/2\kappa^2$. Likewise, the field $\psi$ is dual to an
operator ${\mathcal O}$ in the CFT. Using the standard holographic
dictionary, this has a scaling dimension $\Delta$ fixed by $m$
\cite{Gubser:Breaking,Hartnoll:Building}; for simplicity, we choose
$q=2$ and $m^2=-2/\ell^2$ with $\Delta=2$ \footnote{We can also consider the alternative quantisation with $\Delta=1$ and we expect
analogous results.}. This corresponds to the
expansion $\psi=z\psi_1+z^2\psi_2+\dots$, with $\psi_1=0$. Analogous
to the previous identifications, the AdS/CFT correspondence allows one
to identify $\psi_1$ as a source for the operator ${\mathcal O}$, and
$\psi_2$ as the expectation value, $\langle {\mathcal O}\rangle\equiv
\psi_2/2\kappa^2$.

As highlighted in Refs.~\cite{Gubser:Breaking,Hartnoll:Building}, the
operator ${\mathcal O}$ corresponds to the superfluid order parameter
of the CFT, and is non-vanishing below $T_c$. In the gravitational
framework, this reflects a change in the classical black hole
solutions of the model (\ref{model}); in the normal state the black
holes have $\psi=0$, whilst in the superfluid state they have ``scalar
hair'' with $\psi\neq 0$. The bosonic order parameter ${\mathcal O}$
is argued to be composed of fermionic bilinears and scalar fields
residing in the 2+1 D CFT \cite{Gubser:2009qm}.
Although more complicated than in BCS theory, it is highly reminiscent
of a pairing field. Whilst a detailed microscopic description of the
CFT and its operator content requires a ``top-down'' approach based on
string theory, we can nonetheless make a great deal of progress
without such considerations owing to universality. We will return to
string theory descriptions in future work.

{\em Gaussian Quantum Quench.}--- We now analyse the far from
equilibrium dynamics of the dual CFT, at finite temperature and charge
density, by numerically constructing time-dependent black hole
solutions for the holographic model (\ref{model}). Details of
our coordinate system and metric are provided in the Supplementary
Material; see also Fig.~\ref{Fig:Setup}. 
Near the AdS boundary at $z=0$ the latter 
have the time-dependent asymptotic expansion
\begin{equation}
  \psi=z\psi_1(t)+z^2\psi_2(t)+\dots,\quad  A_t=\mu(t)-z\rho(t)+\dots
  \label{psiexp}
\end{equation}
The holographic renormalization group allows one to establish the
time-dependent correspondence 
\begin{equation}
\label{eq:vevs}
\langle J_t(t)\rangle =\frac{\rho(t)-\dot\mu(t)}{2\kappa^2}, \quad \langle {\mathcal
  O}(t)\rangle=\frac{\left[\psi_2(t)+2i\mu(t)\psi_1(t)\right]}{2\kappa^2},
\end{equation}
in our space-time coordinates and gauge; in the case where
$\mu=\dot\mu=0$ we recover the previous correspondence.  
We take as our initial state
a superfluid corresponding to a black hole with $\psi\neq 0$
\cite{Hartnoll:Building}, and set the initial temperature to
$T_i=0.5T_c$ for numerical convenience; as we will see, similar
results are also expected for other values of $T_i$. We then apply a
quench of the source field $\psi_1(t)$, conjugate to $\langle{\mathcal
  O}(t)\rangle$.  Specifically, we apply a Gaussian-type quench,
centred on $t=0$, by imposing
\begin{equation}
\psi_1(t)= \bar{\delta} \,e^{-(t/\bar{\tau})^2} ,
\label{pulse}
\end{equation}
where $\bar{\delta}$ and $\bar{\tau}$ characterise the quench strength and
time-scale respectively.  The chemical potential of the initial state,
$\mu_i$, sets the scale for the resulting dynamics and explicitly
breaks conformal invariance. This is nonetheless amenable to a holographic
treatment and we use $\mu_i$ to define dimensionless ${\delta} \equiv \bar{\delta} / \mu_i$ and ${\tau} \equiv \mu_i \bar{\tau}$. For definiteness, we set
${\tau}=0.5$ and will vary ${\delta}$.  
We track the dynamics by solving the
equations of motion of \eqref{model} numerically. As discussed in
the Supplementary Material we choose a gauge for $\mu(t)$ which keeps the initial and final charge densities the same while the quench
injects energy into the system. Then $\mu(t)$ interpolates from the initial value, $\mu_i$, to a final chemical potential, $\mu_f$. We find similar results for other values of ${\tau}$, and also for quenches which do not preserve the equality of initial and final charge density.   
As we shall see, our quench is abrupt compared with
the emergent relaxation time-scale.  

\begin{figure}
\includegraphics[clip,width=3.2in]{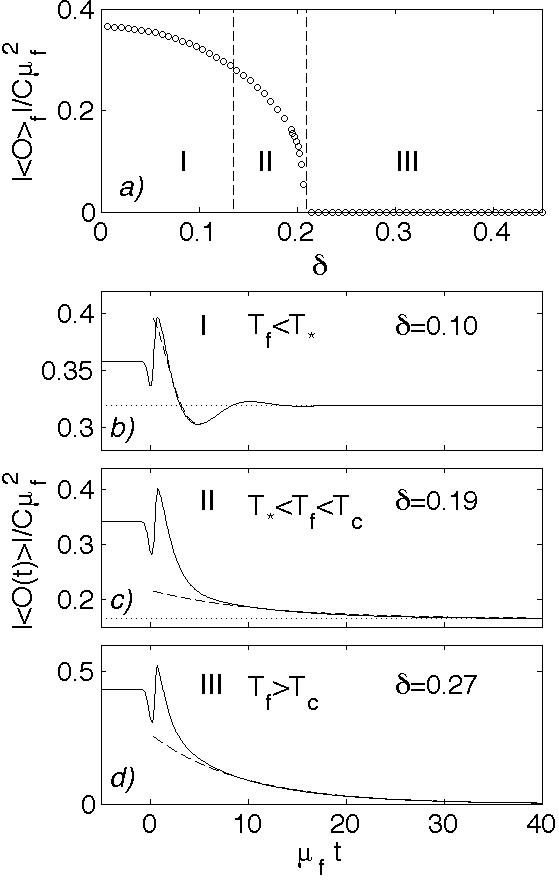}
\vspace*{-0.3cm}
\caption{(a) Dynamical phase diagram of the holographic superfluid
  showing the final order parameter, $|\langle {\mathcal O}\rangle_f|$,
  at late times. We start in the
  superfluid 
with $T_i=0.5T_c$, and monitor the time
  evolution with increasing 
quench strength 
$\delta$. The dynamics
  exhibits three regimes. For the chosen
  parameters the transitions 
  occur at $\delta_\ast\approx 0.14$ and $\delta_c\approx 0.21$.  (b)
  In region I we observe damped oscillations towards $|\langle
  {\mathcal O}\rangle_f|\neq 0$. (c) In II we find a non-oscillatory
  approach towards $|\langle {\mathcal O}\rangle_f|\neq 0$. (d) In III
  we find a non-oscillatory decay towards $|\langle {\mathcal
    O}\rangle_f|=0$. The dashed lines 
in (b), (c), and (d) 
correspond
  to the dominant quasi-normal modes of the final state black holes
  for temperatures $T_f/T_c=0.73,0.95,1.48$ respectively. }
\label{Fig:PD}
\end{figure}

{\em Dynamical Phase Diagram.}--- In Fig.~\ref{Fig:PD} we show the
dynamical phase diagram as a function of
$\delta$.
It displays three regimes of late-time 
behaviour whose asymptotics are governed by the gauge-invariant equation
\begin{equation}\label{vev}
|\langle {\mathcal O}(t)\rangle| \simeq 
|\langle {\mathcal
  O}\rangle_f+{\mathcal A}e^{-i\omega t}|,
\end{equation}
where $\langle {\mathcal O}\rangle_f$ is the final order parameter,
${\mathcal A}$ is an amplitude pre-factor, and $\omega$ is a complex
frequency in the lower half-plane.  In region III it displays
exponential decay towards a vanishing final order parameter $|\langle
{\mathcal O}\rangle_f|=0$, so that for large $\delta$ we exit
the initial superfluid phase completely.  In contrast, in region II it
exhibits non-oscillatory exponential decay with ${\rm Re}(\omega)=0$
towards $|\langle {\mathcal O}\rangle_f|\neq 0$. As we shall see
later, this corresponds to the presence of a gapped ``amplitude'' mode
and a gapless ``phase'' mode in the superfluid phase. However, in
region I it exhibits exponentially damped oscillations with ${\rm
  Re}(\omega)\ne 0$ towards $|\langle {\mathcal O}\rangle_f|\neq 0$,
so that for smaller $\delta$ there is another regime of
dynamics.  For the parameters used in Fig.~\ref{Fig:PD}, the
transition from I to II occurs at a critical value
$\delta_\ast\approx 0.14$, whilst the transition from II to III
occurs at $\delta_c\approx 0.21$.

The behaviour shown in Fig.~\ref{Fig:PD} is
reminiscent of the dynamics of 
a BCS superconductor
\cite{Barankov:Sync}, despite the fact that the holographic superfluid
is strongly coupled, and that the effects of thermal damping are
incorporated.
Indeed, the persistent oscillations
of the integrable BCS Hamiltonian are replaced here by an under-damped
approach towards $|\langle {\mathcal O}\rangle_f|\neq 0$, whilst the
power-law damped BCS oscillations are replaced by an exponentially
damped approach. The transition at $\delta_\ast$ provides a finite
temperature and collision dominated analogue of the collisionless
Landau damping transition \cite{Barankov:Sync}.

\begin{figure}
\includegraphics[clip,width=3.2in]{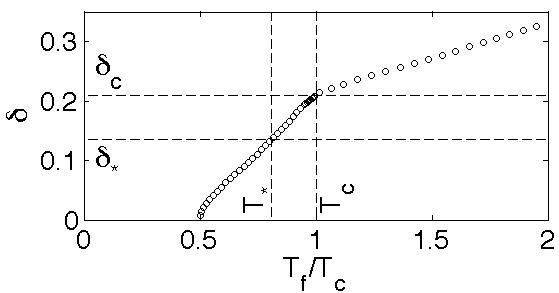}
\vspace*{-0.3cm}
\caption{
  Quench strength $\delta$ versus final state temperature $T_f$ using
  the same initial parameters as in
  Fig.~\ref{Fig:PD}. 
   The dynamical transition at $\delta_\ast\approx 0.14$ 
occurs within the superfluid 
at a
  temperature $T_\ast\approx 0.81 T_c$.}
\label{Fig:Temp}
\end{figure}

It is illuminating to consider the phase diagram as a function of the equilibrium
temperature of the final state black hole, $T_f$. 
Fig.~\ref{Fig:Temp} shows
that $T_f$ increases monotonically with $\delta$, as expected.  
Replotting the data in Fig.~\ref{Fig:PD} against
$T_f$ we obtain the equilibrium phase diagram of the holographic superfluid
\cite{Hartnoll:Building}, with 
the transition from II to III being associated with $T_c$,
and the transition from I to II associated with an emergent temperature scale 
$T_\ast\approx 0.81 T_c$, determined by $\delta_\ast$.

\begin{figure}
  \includegraphics[clip,width=3.2in]{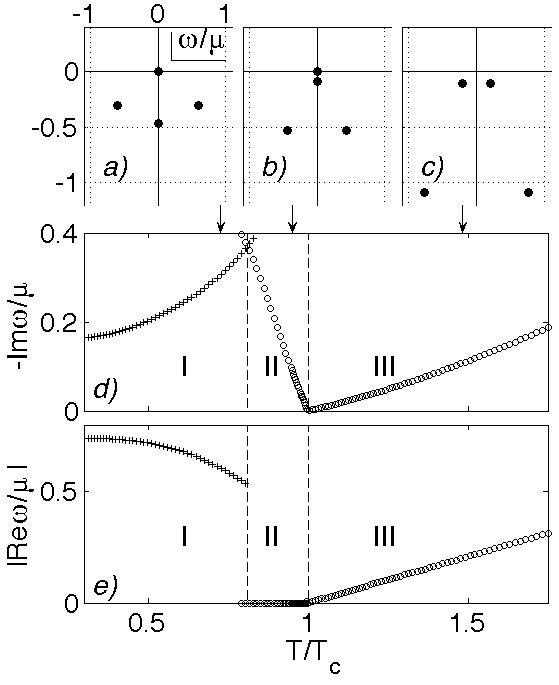}
\vspace*{-0.3cm}
  \caption{Evolution of the QNM frequencies with temperature. (a)
    {$T=0.73T_c$}. (b) {$T=0.95T_c$}. (c) {$T=1.48T_c$}. Time reversal
    invariance corresponds to $\omega\rightarrow-\omega^*$. (d) and
    (e) show the imaginary and real parts of the dominant
    QNMs. i.e. the QNMs closest to the real axis. The results show
    three regimes of dynamics, in quantitative agreement with
    Fig.~\ref{Fig:PD}.}
\label{Fig:QNM}
\end{figure}

{\em Quasi-Normal Modes.}--- 
To gain insight into the three regimes of collective dynamics and the
temperature $T_\ast$, we examine the late time asymptotics in
more detail.
As $t\rightarrow\infty$, the
dynamics  
is described by the quasi-normal modes (QNMs) of the late time black holes.
Each QNM describes an approach to equilibrium in linear perturbation theory with time dependence $e^{-i \omega t}$. Those that dominate the late time dynamics have complex frequency $\omega$ 
closest to the real axis
and give rise to the behaviour in Eq.~(\ref{vev}).
As outlined in the
Supplementary Material, we have calculated the 
homogeneous isotropic QNMs
both for the normal state black holes
(see Eq.~ (2) of the Supplementary Material), and for the superfluid black holes of
Ref.~\cite{Hartnoll:Building}. 
This generalises the analysis of \cite{Amado:2009ts} who calculated the QNMs (also for non-zero momentum) in a probe approximation.
For our purposes, we need to go beyond the probe approximation and include back-reaction and the trajectories of the dominant QNMs in the complex $\omega$ plane
are depicted in Fig.~\ref{Fig:QNM}. 

Typically, the real parts of the dominant QNM frequencies correspond
to oscillations, and the imaginary parts to damping.  However, as
shown in Fig.~\ref{Fig:QNM}(c), for 
$T>T_c$, the QNMs
for the normal state black hole have two complex frequencies that are
closest to the real axis. Nonetheless, substitution into
Eq.~(\ref{vev}) with $\langle{\cal O}_f\rangle=0$ yields the damped
non-oscillatory behaviour found in region III of Fig.~\ref{Fig:PD}.
As the temperature is lowered, these dominant poles migrate upwards in
the complex $\omega$ plane and at the superfluid transition
temperature, $T_c$, they coincide at $\omega=0$. This corresponds 
to 
spontaneous ${\rm U}(1)$ symmetry breaking with the
appearance of a Goldstone mode. Below $T_c$, one of these modes, the
``amplitude'' mode, travels down the imaginary axis, consistent with
time-reversal invariance under $\omega\rightarrow -\omega^\ast$,
whilst the Goldstone ``phase'' mode remains pinned at $\omega=0$; see
Fig.~\ref{Fig:QNM}(b).  The amplitude mode describes the damped
approach to a finite order parameter 
in region II of Fig.~\ref{Fig:PD}; the Goldstone mode does not affect
the dynamics in the homogeneous and isotropic context, although it
does lead to a hydrodynamic mode at non-zero spatial momentum.  As the
temperature is lowered, the subdominant poles also ascend in the
complex plane.  At the dynamical transition temperature $T_\ast$,
the damping rate of the descending amplitude mode coincides with that
of the ascending subdominant complex poles.  For the chosen parameters
this occurs at $T_\ast\approx 0.81 T_c$, in agreement with the
nonlinear analysis. 
Below
$T_\ast$, the previously subdominant poles now become dominant, as
shown in Fig.~\ref{Fig:QNM}(a).  The dynamics corresponds to a damped
oscillatory approach to a finite order parameter as found in region I.
In addition to this change in dynamics at $T_\ast$, one may also
extract the variation of the emergent timescales as a function of
temperature. As shown in Figs.~\ref{Fig:QNM}(d) and (e), there are
three regimes.  Moreover, the extracted timescales are in quantitative
agreement with the late time behaviour of the nonlinear analysis, as
indicated by the dashed lines in Figs.~\ref{Fig:PD}(b)-(d). 
The linear response analysis provides an excellent
description over a broad time interval.

{\em Dynamics of Symmetry Breaking.}--- The main results on the late
time behaviour of the quenched holographic superfluids, captured in
Figs.~\ref{Fig:PD}-\ref{Fig:QNM}, have a more universal
applicability. Recall that the location of the QNMs of the black holes
presented in Fig.~\ref{Fig:QNM}, correspond to the location of poles
of the retarded Green's function for the operator ${\cal O}$ in the
dual theory \cite{Berti:2009kk}. Thus, the late time behaviour is
equivalently described by the poles of the retarded Green's function
that are closest to the real axis.  A key point is that the pole
structure in Fig.~\ref{Fig:QNM} is the {\it generic} behaviour for an
isotropic and homogeneous system with time-reversal invariance under
$\omega\rightarrow -\omega^*$, which can spontaneously break a
continuous global symmetry including the presence of the Goldstone
mode at the origin and secondary quasiparticle excitations. The value
of $T_\ast$, if it exists, will be given by the temperature at which
the value of ${\rm Im}(\omega)$ for the pole on the imaginary axis,
and those poles off the imaginary axis and closest to the real axis,
coincide.  At temperatures less than $T_\ast$ there could also be
additional dynamical temperature scales.  For a local symmetry we also
expect analogous phenomenology with the Goldstone mode replaced by the
longitudinal mode of the massive vector. It would be interesting to
compute the pole structure in other models \footnote{See
  \cite{Podolsky:2012pv} for a recent calculation of the spectral
  properties of the ${\rm O}(N)$ model at zero temperature.},
including non-conformal geometries, and to
explore the ramifications in experiment. Recent experiments using cold
atomic gases \cite{Endres:Higgs} suggest the possibility of investigating
the evolution of the excitation spectrum.

{\em Acknowledgments.}--- We thank P. Chesler, A. Green, S. Hartnoll,
P. Figueras, K. Landsteiner,  L. Lehner, R. Myers, S. Sachdev, K. Schalm and
D. Tong for  
discussions.  
We thank GGI, KITP, Leiden and PI for hospitality and
acknowledge EPSRC grant EP/E018130/1 and NSF grant PHY05-51164.

\bibliography{hologsuperbib}

\section{SUPPLEMENTARY MATERIAL}

{\em Nonlinear Dynamics.}--- To investigate the dynamics of the model
we begin by introducing space-time coordinates and the
metric. Assuming spatial homogeneity and isotropy of the CFT under
time evolution, the most general metric is
\begin{eqnarray}\label{ansatz}
ds^2 =z^{-2} \left[ -F \,dt^2 - 2 \,dt dz + S^2(dx_1^2+dx_2^2) \right],
\end{eqnarray}
where $(t,x_1,x_2)$ are common to both the boundary and bulk theories,
and 
$z$ specifies the additional direction in the bulk
space-time. This corresponds to ingoing Eddington--Finkelstein
coordinates, where the asymptotic AdS boundary is located at $z=0$;
see Fig.~1 of the letter. Here $F(t,z)$ and $S(t,z)$ depend only on $t$
and $z$.  Likewise $\psi=\psi(t,z)$,
$A_t=A_t(t,z)$ and the 
spatial components of $A_a$ are set to zero. 
Thus, the dynamics is specified by five (real) functions of $t$ and $z$.

We next recall
that in equilibrium, the CFT at finite
temperature and charge density is described by an electrically charged
static black hole.
The high temperature unbroken phase of the 
CFT is described by 
the AdS-Reissner--Nordstr\"om (AdS-RN) black hole
\cite{Gubser:Breaking,Hartnoll:Building}
with 
\begin{align}\label{adsrn}
F=1-2Mz^3+(\rho^2/4)z^4,\quad S=1,\quad A_t=\mu-\rho z,
\end{align}
and $\psi=0$. Here $\mu$ and $\rho$ are the chemical potential and the
charge density of the dual CFT,
with
$\langle J_t\rangle\equiv \rho/2\kappa^2$. Likewise, the mass $M$ is
proportional to the energy density of the CFT, and the Hawking
temperature, $T_{\rm H}(\rho,M)$, corresponds to the temperature $T$
of the CFT.  At $T_c\approx 0.090\mu$ the AdS-RN black hole becomes
unstable and the CFT is described by a new family of black hole
solutions with $\psi\neq 0$
\cite{Hartnoll:Building}.  
Asymptotically
close to the AdS boundary $\psi$ has the coordinate expansion
$\psi=z\psi_1+z^2\psi_2+...$ with $\psi_1=0$. 
Analogous to the identifications
following Eq.~(\ref{adsrn}), the AdS/CFT
correspondence allows one to identify $\psi_1$ as a source for the
superfluid order parameter in the dual CFT, and $\psi_2$ as the
expectation value, $\langle {\mathcal O}\rangle\equiv
\psi_2/2\kappa^2$. Hence, these black holes with $\psi\neq 0$ describe
a superfluid phase in which the global ${\rm U}(1)$ symmetry is
spontaneously broken. 

To study the response to the quench given
in 
Eq.~(\ref{pulse})
we solve the equations of motion 
numerically, using the metric described near
Eq.~(\ref{ansatz})
(see also 
\cite{Murata:Noneq}).  The asymptotic
boundary is located at $z =0$ and writing $\psi(t,z) =z[\psi_1(t) +
\tilde{\psi}(t,z)]$, $a(t,z) = \mu(t) + \tilde{a}(t,z)$, $F(t,z) =1 +
z^2[ -|\psi_1|^2/2+\tilde{F}(t,z)]$ and $S(t,z) = 1 +
z^2[-|\psi_1|^2/4+\tilde{S}(t,z)]$ we can choose a gauge where
$\tilde\psi\sim \psi_2(t)z$, $\tilde a\sim \rho(t)z$, and $\tilde
F,\tilde S$ also vanish linearly as $z \to 0$. Notice that the
residual coordinate freedom $1/z\to1/z+f(t)$ is fixed by these
asymptotics.  In contrast to 
\cite{Murata:Noneq}, we adopt a gauge for
$\mu$ so that ${\rm Im}(\psi_2-D\psi_1)=0$, where
$D=\partial_t-2i\mu$.  One can show that the equations imply the
boundary charge conservation equation, given by $\dot\rho=-4{\rm
  Im}[\psi_1^*(\psi_2-D\psi_1)]$, and also a (sourced) energy
conservation equation.
Our choice of quench, with $\psi_1$ real as in 
Eq.~(\ref{pulse}),
ensures that the initial and final charge densities are the same.
Specifically, this can be seen by considering 
Eq.~(\eqref{eq:vevs})
and observing that
we have $\dot\rho=0$ and also $\dot \mu=0$ at $t=\pm\infty$.
With $S,F,A_t$ and the complex scalar $\psi$ we have 5 real quantities
to evolve as functions of the coordinates $t, z$. There are 8 (real)
equations of motion from the metric, vector and scalar equations, 5 of
which we use as evolution equations for the variables $\tilde{\psi},
\tilde{a}, \tilde{F}, \tilde{S}$, and the remaining 3 are constraints.
The 5 equations have principal parts, $\partial^2_{tz} \psi -
\frac{1}{2} F \partial^2_{z} \tilde{\psi}=0$, $\partial^2_{tz}
\tilde{a}=0$, $\partial^2_{tz} \tilde{S}=0$, and $F \partial^2_{z}
\tilde{S} + S \partial^2_z \tilde{F}=0$, where the latter is
elliptic. Provided the initial data satisfies the 3 constraints then
two of them, with principal parts $\partial^2_{z} \tilde{S} =0$ and
$\partial^2_{z} \tilde{a}=0$, are automatically satisfied at later
times. The one remaining constraint, with principal part
$\partial^2_{t} \tilde{S}=0$, must be imposed at the boundary in
addition to the 5 equations of motion in the bulk. Physically, this
corresponds to (sourced) boundary stress energy conservation, and
provides the additional data required for the elliptic evolution
equation. Note that (sourced) current conservation results from the
evolution equations and is not imposed separately.

To solve the equations of motion numerically we use a Chebyshev pseudo
spectral representation in $z\in [ 0,1 ]$, where the initial 
horizon is located at $z=1$, and use an implicit Crank-Nicholson
finite difference scheme in $t$. Given a slice at constant $t$ we
advance to the next slice by solving for $\tilde{\psi}, \tilde{a},
\tilde{F}, \tilde{S}$ from the 5 evolution equations together with the
boundary constraint and gauge condition for $\mu$; note that we are
not required to impose any boundary condition at the innermost point
$z =1$ since this is inside the event horizon. The method outlined
above is very robust, and by virtue of the spectral representation in
$z$ it allows very accurate extraction of boundary quantities. We find
that relatively modest grid sizes with 20 points in $z$ already give
reliable results. The data presented in this paper is for $40$ points,
where convergence testing indicates that the errors will be less than
percent level in all the plotted quantities. We construct static
superfluid black hole solutions of 
\cite{Hartnoll:Building} 
for the
initial data by solving the usual ordinary differential equation
shooting problem in our gauge. 
We then find that we can stably evolve 
up to times $t \sim 60\, \mu_i^{-1}$, to extract the data presented here. If one runs
for too long then we sometimes encounter the black hole singularity.
In order to evolve further one must implement singularity excision
which we leave for future work. Our present results are in full quantitative
  agreement with the QNM analysis described in the text and below. 

Finally, we note that the event horizon, depicted in figure 1, is obtained by tracing back a null ray from the final equilibrium black hole horizon.

{\em Quasi-Normal Modes (QNMs).}--- The QNMs of the equilibrium black
holes of 
Eq.~(\ref{model})
are linearised perturbations with ingoing
boundary conditions at the black hole horizon and normalisable
boundary conditions at the asymptotic AdS boundary; for a review see
 \cite{Berti:2009kk}. For the purposes of this paper we only
consider the zero momentum sector of the QNM spectra.
Furthermore, to determine the late dynamics of the charged scalar
field, $\psi$, we only analyse the sector involving $\psi$. For the
AdS-RN black hole 
(\ref{adsrn})
we find a second order ODE for
$\psi$. For the superfluid black hole of 
\cite{Hartnoll:Building} which has $\psi\ne 0$, we can use the gauge freedom, arising from
diffeomorphisms and local ${\rm U}(1)$ transformations of the
background, to reduce the problem to two second order ODEs for two
gauge invariant variables involving $\psi$, $A$ and the metric.  In
both cases we then use Chebyshev pseudo-spectral differencing to cast
the linear perturbation equations into the form, ${\rm M}(\omega;
\lambda) \mathbf{v}=0$. The matrix ${\rm M}$ depends on the complex
frequency $\omega$ and the background parameters $\lambda$, such as
the temperature and chemical potential.  The vector $\mathbf{v}$
consists of the two gauge-invariant variables evaluated at the grid
points. The QNM frequencies are then determined by the condition
$|{\rm M}(\omega;\lambda)|=0$.

\end{document}